\documentclass[12pt]{article}
\usepackage{newtxtext,newtxmath}
\usepackage[letterpaper,margin=1in]{geometry}

\usepackage{scicite}
\usepackage{graphicx}
\usepackage{times}
\usepackage{comment}

\renewenvironment{abstract}
	{\quotation}
	{\endquotation}

\makeatletter
\renewcommand{\fnum@figure}{\textbf{Figure \thefigure}}
\renewcommand{\fnum@table}{\textbf{Table \thetable}}
\makeatother

\usepackage{scicite}

\usepackage{url}

\usepackage{xcolor}

\def\scititle{
	Ultrafast momentum-resolved visualization of the interplay between phonon-mediated scattering and plasmons in graphite
}
\title{\bfseries \boldmath \scititle}

\author{
	Francesco Barantani$^{1,2,\dag}$, 
        Rémi Claude$^{1,3,\dag}$, 
        Fadil Iyikanat$^4$,
        Ivan Madan$^1$, \and
        Alexey Sapozhnik$^1$, 
        Michele Puppin$^{1,3}$, 
        Bruce Weaver$^{1}$,
        Thomas LaGrange$^1$,\and
        F.~Javier Garc\'{\i}a~de~Abajo$^{4,5}$,
        Fabrizio Carbone$^{1,\ast}$\and
	\small$^{1}$Institute of Physics, École Polytechnique Fédérale de Lausanne, Lausanne, 1015, Switzerland\and
        \small$^{2}$Department of Quantum Matter Physics, University of Geneva, Geneva, 1211, Switzerland\and
	\small$^{3}$Lausanne Centre for Ultrafast Science, École Polytechnique Fédérale de Lausanne,\and \small  Lausanne, 1015, Switzerland\and
        \small$^{4}$ {ICFO-Institut de Ciencies Fotoniques}, { The Barcelona Institute of Science and Technology},\and \small{{Castelldefels (Barcelona)}, {08860}, {Spain}}\and
        \small$^{5}${{ICREA}, {Instituci\'o Catalana de Recerca i Estudis Avan\c{c}ats}, {{Barcelona}, {08010}, {Spain}}}\and
	\small$^\ast$Corresponding author. Email: fabrizio.carbone@epfl.ch\and
	\small$^\dagger$These authors contributed equally to this work.
}

\date{}

\begin{document} 

\baselineskip24pt

\maketitle 

\begin{abstract}\bfseries \boldmath
     Scattering between individual charges and collective modes in materials governs fundamental phenomena such as electrical resistance, energy dissipation, switching between different phases, and ordering. The study of such scattering requires a simultaneous access to the ultrafast momentum-resolved dynamics of single-particle and collective excitations, which remains as an experimental challenge. Here, we demonstrate time- and momentum-resolved electron energy-loss spectroscopy, and apply it to graphite showing that large ($\Delta q\simeq$1.2~\AA$^{-1}$) photoexcited electron-hole (e-h) pockets in the band structure induce a renormalization of the collective in-plane and bulk plasmons that can be described quantitatively by invoking intra- and inter-valley scattering processes mediated by $E_{2g}$ and $A_1'$ phonon modes, which we directly observe by ultrafast electron diffraction and identify {\it via} {\it ab initio} calculations. Conversely, the photoexcitation of smaller e-h pockets ($\Delta q\simeq$0.7~\AA$^{-1}$) close to the K point of graphite results in the renormalization of in-plane plasmons, which can only be partially explained by phonon-mediated scattering and thermal expansion. Our results show the importance of combining momentum- and time-resolved information to elucidate microscopic details associated with electronic scattering processes.
\end{abstract}

\section*{Introduction}

Understanding and harnessing the properties of quantum materials is key to implementing novel functionalities in future devices. In the realm of two-dimensional materials, a vast range of engineered materials are accessible through approaches that include assembling layered heterostructures~\cite{Novoselov2016}, controlling the interlayer twist angle~\cite{Cao2018}, in-plane sliding~\cite{ViznerStern2021}, or exploiting the valley degree of freedom~\cite{Mak2012}. 
The latter leverages on manipulating carriers present at specific valleys in the electronic band structure (i.e., minima of the conduction band or maxima of the valence band) close to the Fermi level~\cite{Schaibley2016}.
Graphene is a prototypical two-dimensional system, presenting inequivalent Dirac points where valley polarization is observed~\cite{Rycerz2007}, and thus serving as an ideal model for exploring and investigating valleytronics scenarios.

Most of the studies on valley polarization were performed by applying optical techniques and focused on emergent excitons with a specific valley character~\cite{Schaibley2016}. Deeper insight into intervalley scattering can be obtained by gathering time-resolved information to extract, for example, specific relaxation time scales given by different scattering channels.
However, due to the small momentum carried by visible photons, optical measurements are limited to the exploration of scattering processes involving collective modes close to the $\Gamma$ point. This severely limits the possibility of studying processes at finite momenta, such as those mediated by zone-boundary phonons~\cite{Stern2018} or involving dark excitons~\cite{Madeo2020}.

In addition, finite momentum spectroscopies resolved on ultrafast time scales are mostly restricted to study the single-particle response. This is the case of time-resolved angle-resolved photoemission spectroscopy (tr-ARPES), where the effect of bosonic interactions is averaged over momentum transfers within the Brillouin zone (BZ). Only recently, X-ray free-electron lasers (X-FELs) have enabled the realization of direct finite-momentum spectroscopy of the transient collective response in quantum materials~\cite{Dean2016, Mitrano2019, Wandel2022}, which complements the information acquired on the single-particle response.

\begin{figure*}[ht!]
    \centering
    \includegraphics[width =0.9\linewidth]{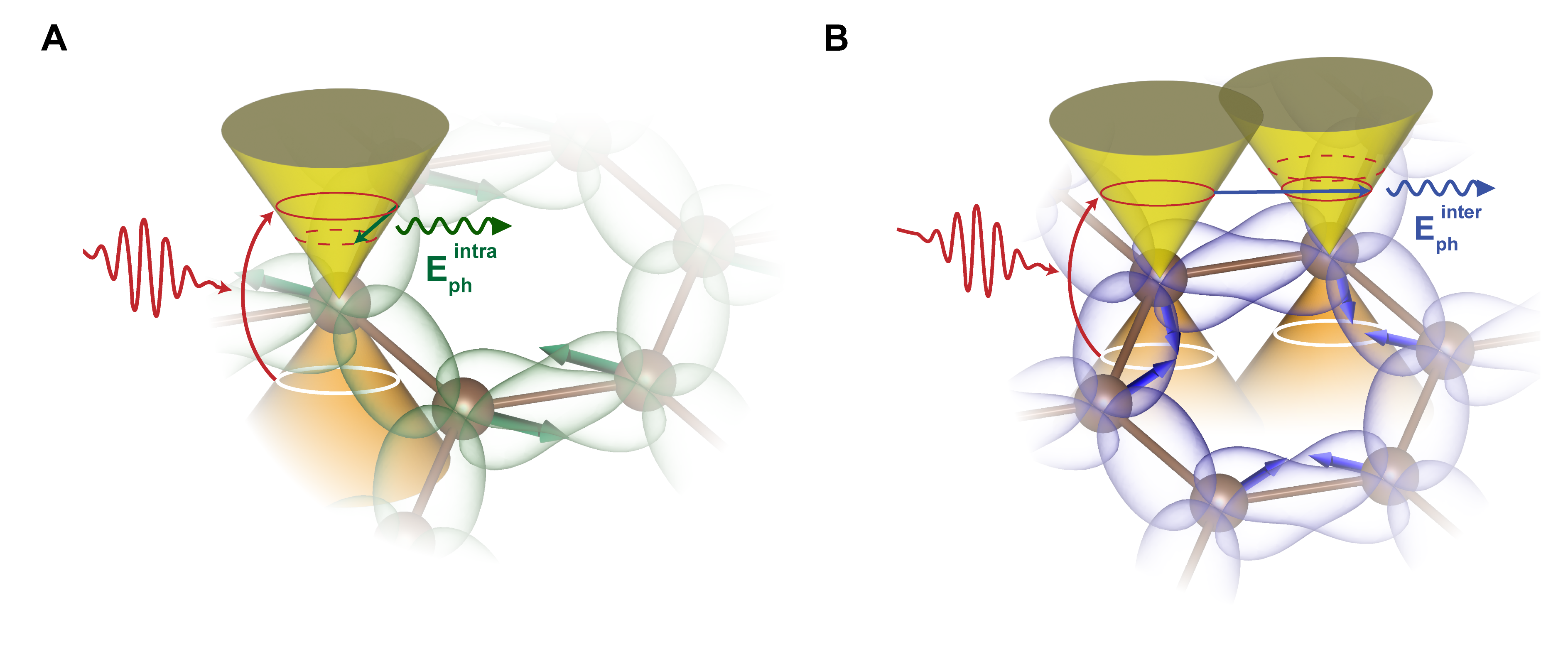}
    \caption{\textbf{Carrier scattering in photoexcited graphite.} Sketch of intravalley (\textbf{A}) and intervalley (\textbf{B}) scattering processes occurring in graphite. An ultrafast laser pulse photoexcites the electrons, which can relax through the so-called strongly coupled optical phonons. Atomic displacements are indicated by green and glue arrows for the $E_{2g}$ (intravalley) and $A_1'$ (intervalley) phonon modes. To pictorially show the spatial distribution of electronic charges reacting to the above-mentioned scattering processes, we plot the charge density associated with molecular orbitals of the $\pi$ bands crossing the Fermi level.}
    \label{fig:fig1}
\end{figure*}

Here, we use a combination of momentum-resolved techniques to investigate the ultrafast scattering processes occurring across different valleys in graphite. First, through high signal-to-noise-ratio ultrafast electron diffraction (UED) measurements, we obtain the out-of-equilibrium population of the phonons involved in the different scattering mechanisms ensuing laser photoexcitation. Subsequently, we carry out a new form of time- and momentum-resolved electron energy-loss spectroscopy (tr-q-EELS)
in an ultrafast transmission electron microscope (UTEM) to visualize the impact that such intra- and inter-valley scattering processes have on collective electronic modes (plasmons). 
After photoexcitation close to the Dirac points, hot electrons induce an out-of-equilibrium phonon population~\cite{Moos2001, Kampfrath2005}, which produces scattering within the same valley --thus interacting with $E_{2g}$ phonons close to the $\Gamma$ point-- or across different valleys --involving $A_1'$ phonons with momentum close to $K$. These two relaxation mechanisms are illustrated in panels A and B of Figure~\ref{fig:fig1}, respectively, together with a sketch of the charge density distribution in graphite.

In graphite, one can readily harness the excitation photon energy to tune the size of photoexcited electron-hole (e-h) pockets in the electronic momentum distribution. We utilize near-infrared (NIR, 1.55~eV) and visible (3.1~eV) pulses to respectively photoexcite small ($\Delta q \le$0.7~\AA$^{-1}$) and large ($\Delta q \ge$1.2~\AA$^{-1}$, more isotropic) e-h pockets around the $K$ points. We then monitor the impact of phonon-mediated charge-carrier relaxation through intra- and inter-valley scattering processes, first by measuring the ultrafast diffused diffraction intensity and secondly by tracking the evolution of plasmonic collective modes by tr-q-EELS. 

\noindent
The temporal evolution of both the in-plane $\pi$ and the bulk $\sigma$+$\pi$ plasmons that we observe for 3.1~eV photoexcitation can be quantitatively modeled by {\it ab initio} calculations, considering a large out-of-equilibrium population of $E_{2g}$ phonons mediating the intervalley scattering processes. Diffuse scattering measurements confirm this scenario by revealing an intense signature of phonons near the $\Gamma$ point, directly related to electronic relaxation of large e-h photoexcited pockets. 
Instead, for the smaller NIR-excited pockets, we need to also consider the out-of-plane thermal expansion produced by an elevation in lattice temperature in order to qualitatively reproduce the experimental loss function dynamics, while we cannot fully explain the $\pi$ plasmon dynamics.

\noindent
The $\pi$ plasmon, which in previous studies was buried in the elastic peak and so impossible to observe~\cite{Carbone2009}, undergoes different dynamics for the two pump energies. This observation can be rationalized by considering that the $\pi$ plasmon results from the $\pi$ and $\pi^{\star}$ bands that cross the Fermi surface and can experience additional screening due to the increases in phonon population and electronic temperature. Furthermore, previous studies~\cite{Pagliara2011} suggested that the same $\pi$ and $\pi^{\star}$ bands can show a strong electronic renormalization when resonantly excited. 

\begin{figure*}[ht]
    \centering
   \includegraphics[width=\linewidth]{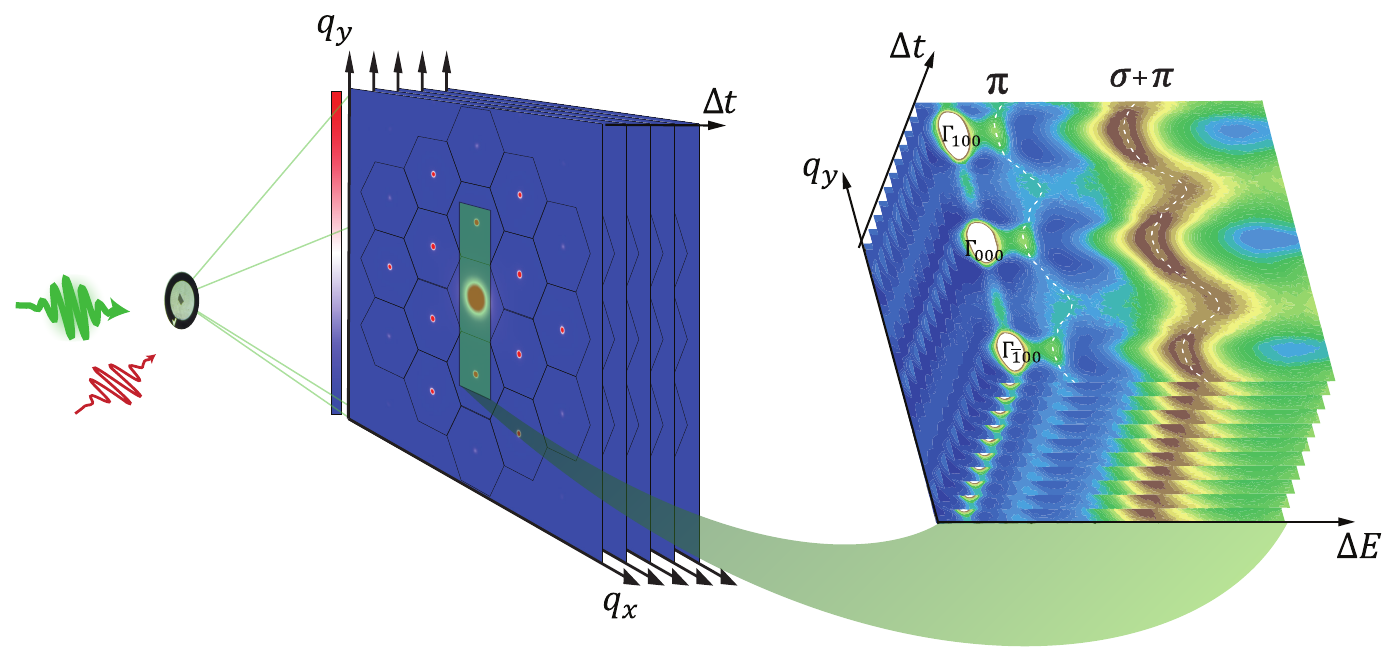}
    \caption{\textbf{Illustration of the experimental approach employed for momentum-, time- and energy-resolved measurements.} An ultrafast electron diffraction (UED) experiment (left) measures the diffraction patterns resulting upon scattering of an electron pulse (green) following photoexcitation by an optical pulse (red). Time- and momentum-resolved electron energy-loss spectroscopy (tr-q-EELS) data (right) arises from the energy dispersion of the scattered electrons selected by a slit oriented along the $\Gamma \to M$ direction (green rectangle in central plot). The right panel shows measured tr-q-EELS data at different time delays.}
    \label{fig:exp_fig}
\end{figure*}

\section*{Results}
Because of its profound similarities with graphene, and thanks to the relatively well-understood behavior under equilibrium conditions, out-of-equilibrium graphite has been the subject of extensive investigations throughout the last decades. Optical~\cite{Moos2001, Kampfrath2005, Breusing2009, Pagliara2011}, structural~\cite{Carbone2008, Chatelain2014}, and electronic measurements at high energies~\cite{Carbone2009, Carbone2009a, vanderVeen2015} provided insight into the characteristic relaxation channels on different timescales, while ultrafast diffused electron scattering~\cite{Stern2018, deCotret2019} and tr-ARPES~\cite{Stange2015, Yang2017, Na2019, Beyer2023} experiments revealed the important role played by strongly coupled optical phonons in the scattering processes. The plasmonic properties of graphite have also attracted attention due to the long lifetimes observed in the low-energy plasmons of graphene and their potential for applications~\cite{Koppens2011, Grigorenko2012}. However, a direct observation of the interplay between single-particle excitations and collective electronic and structural modes is lacking to date due to the challenged involved in obtaining microscopic information that is simultanenously momentum- and time-resolved.

\noindent
Figure~\ref{fig:exp_fig} illustrates our combined experimental approaches --UED, on the left, and tr-q-EELS, on the right--, both or which involving an electron pulse that is transmitted through a graphite flake excited by an ultrafast laser pulse. UED provides access to the response in reciprocal space: incoherent lattice motion, Debye-Waller effect, and phonon population imprinted on the Bragg and diffused electrons, allowing for the measurement of the time-resolved dynamics of low-energy collective modes. Tr-q-EELS tracks the spectroscopic response of the inelastically scattered electrons, yielding the energy and momentum-resolved loss function of the material. In this study, we place a slit in reciprocal space to select a specific momentum region oriented along the desired direction, and we collect energy-loss spectra using an electron spectrometer combined with an UTEM. This results in a direct measurement of the electronic collective mode dispersion as a function of transferred momentum.

\begin{figure*}
    \centering
   \includegraphics[width=\linewidth]{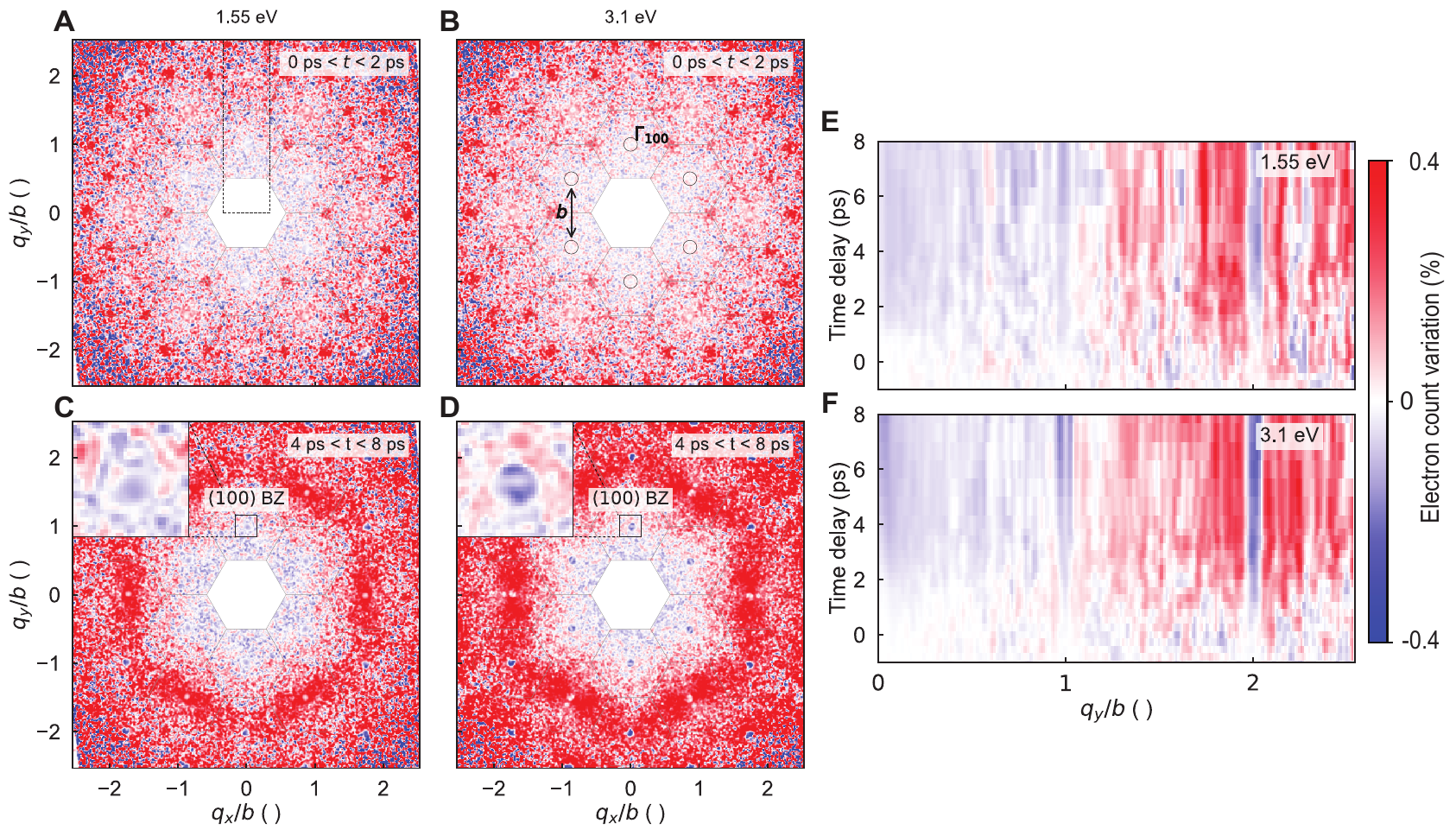}
    \caption{\textbf{Ultrafast diffuse response measured by ultrafast electron diffraction.} (\textbf{A}--\textbf{D}) UED maps of graphite after excitation with 1.55~eV (A, C) and 3.1~eV (B, D) photon energy, normalized to the unpumped diffraction patterns. Panels (A) and (B) show diffraction patterns averaged over short time delays (0.1 to 2~ps), while (C) and (D) are averaged over longer
    time delays (4 to 8 ps). Insets are zooms of a first-order Bragg peak. (\textbf{E}, \textbf{F}) Diffraction patterns binned along the $\Gamma\to M$ direction (within the dashed rectangle in (A)) as a function of time delay after the photoexcitation by 1.55~eV and 3.1~eV pump pulses, respectively. The color scale is shared among the six maps. In panels (E, F), the wave vector is normalized to the reciprocal lattice unit $b=2.95$~{\r{A}} (see double arrow in (A)).}
    \label{fig:UED_graphite}
\end{figure*}

\subsection*{Ultrafast electron diffraction}

We present ultrafast electron diffraction measurements performed on a 50 nm-thick natural graphite flake with a shot-to-shot single electron detector acquisition (see SI). We  observe the photo-induced effect by normalizing the photoexcited diffraction pattern by the one without excitation.
Figures~\ref{fig:UED_graphite}A-D show the diffused scattering response averaged over two different time-delay intervals (0.1--2~ps and 4--8~ps) for the two pump energies (1.5 and 3.1 eV) at the same absorbed fluence of 0.8 mJ/cm$^2$. Each Bragg peak corresponds to the center of each BZ, represented by hexagons. At early stages (0.1--2~ps delay), we observe a strong intensity increase at the $K$ points of the BZ (Figure~\ref{fig:UED_graphite}A, B) --an effect produced by the increase in the population of $A_1'$ phonons~\cite{Stern2018}. At later larger delays (4--8~ps), we observe an increase in intensity between the second-order Bragg peaks and the $M$ points of the reduced BZ, which we attribute to excitations of transversal acoustic phonons with a momentum spanning the $\Gamma\to M$ direction. Due to the low excitation fluence, the incoherent thermal motion of the lattice is small as confirmed by the weak Debye-Waller effect at the Bragg peak (see Figure~S4A). \\ 
The panels C and D of Figure~\ref{fig:UED_graphite}, show a clear difference in the $\Gamma_{100}$ responses at the two pump energies, as highlighted in the insets. To better understand this behavior, we integrate the diffraction patterns along the $q_x$ axis (in the region illustrated by the dashed rectangle in A) and plot the dynamics along the $\Gamma \to M$ direction. The results are represented in Figure~\ref{fig:UED_graphite}E, F, where we observe a decrease in intensity at the Bragg peak positions for both photoexcitation energies. In contrast to what happens with the 1.55~eV pump, the 3.1~eV response presents an increase in intensity at the center of the $\Gamma_{100}$ point compensating the Debye-Waller attenuation after 3~ps, which can be resolved as a double negative feature around the peak. This effect suggests an increase in phonon population at small momenta (i.e., $E_{2g}$ phonons). In addition, we find that the $\Gamma\to M$ direction exhibits a remarkable difference between the two photon energies, and therefore, we perform tr-q-EELS experiments for momentum resolved along that direction to investigate how the low-energy phonon modes impact the high-energy electronic response. 

\subsection*{Time- and momentum-resolved EELS: experiments and simulations}

Previous investigations of graphite by ultrafast EELS were limited to the bulk $\sigma+\pi$ plasmon dynamics~\cite{Carbone2009} or focused on core-loss peaks to extract the evolution of the local density of states~\cite{vanderVeen2015}. Our approach overcomes such limitations and allows us to also observe the $\pi$ plasmon, located at around
7~eV and directly originatingfrom interband transitions between the $\pi$ and $\pi^*$ electronic bands. 
\noindent
As depicted in Figure~\ref{fig:exp_fig} and highlighted in the SI, we identify the $\pi$ plasmon in the EELS spectra at low energies and the $\sigma+\pi$ plasmon at around 30~eV. Following the dispersion of both plasmons up to the zone boundary, we notice a typical quadratic dispersion consistent with previous literature~\cite{Liou2015}. However, the limited energy resolution of our ultrafast electron microscope does not allow us to directly resolve phonon losses, which are buried within the zero-loss peak.

\noindent
Most of the previous time-resolved EELS studies related to the photon-induced near-field electron microscopy (PINEM) technique~\cite{Barwick2009}, which takes place when laser and electron pulses arrive together on the sample and the latter interacts with the near-field induces near material surfaces. The electron spectrum broadens and exhibits energy lobes equispaced by the photon energy. Even if we adjust the light polarization to minimize such a PINEM effect in our studies, it still emerges in the EELS maps. The very fast (time delay$\le 500$ fs) dynamics is therefore polluted by PINEM interactions in our study, and consequently, we discard it in our discussion.

\noindent
In the dynamics captured by tr-q-EELS, we expect the $\pi$ plasmon to be strongly influenced by the laser photoexcitation and the subsequent scattering processes. On the other hand, the $\sigma+\pi$ plasmon at higher energies is known to be strongly affected by the interlayer spacing\cite{Marinopoulos2004}. 
In our experiments, we illuminate a flake of highly-ordered pyrolytic graphite (HOPG) flake oriented along the [001] zone-axis in analogous conditions to the UED measurements, with 1.55~eV and 3.1~eV ultrafast pulses.  We compare the variation of the EELS spectra as a function of momentum, along the $\Gamma\to M$ high-symmetry direction as motivated before. 

\noindent
In the analysis of UED experiments, we integrate the spectra before excitation by optical pulses to represent the equilibrium situation, and the spectra for relatively long time delays averaged between 2 and 8~ps to study the photoexcited specimen. In this way, we avoid both spurious PINEM effects and fast out-of-equilibrium electronic contributions. We further normalize the the excited spectra in the energy-momentum-resolved maps to the unexcited response and plot the observed variation relative to the equilibrium specimen.

\begin{figure*}[ht]
    \centering
    \includegraphics[width=\linewidth]{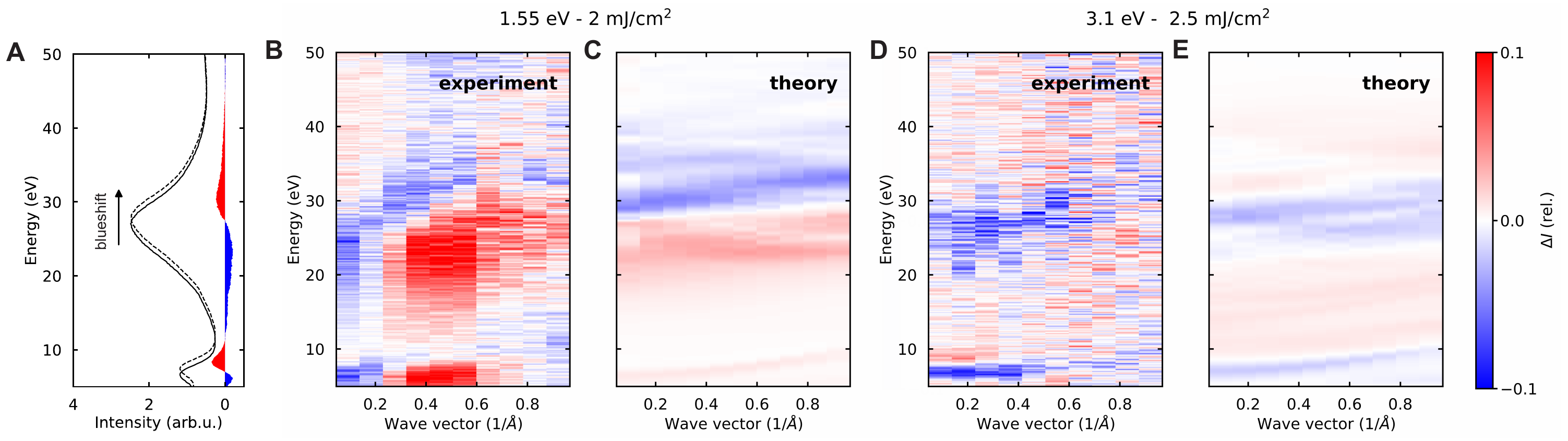}
    \caption{\textbf{Experimental and theoretical variation of the EELS response in photoexcited graphite.} (\textbf{A}) Graphite EELS response at the $\Gamma_{000}$ point. We show a spectrum acquired from unperturbed graphite (solid curve), a blue-shifted version of this spectrum (dashed curve), and the difference between them (filled blue/red areas for negative/positive values). (\textbf{B-E}) Wave vector dependence of the pump-induced change in EELS. (\textbf{B}, \textbf{D}) Maps measured along the $\Gamma\to M$ direction for 1.55 eV pump with 2~mJ/cm$^2$ fluence (B) and 3.1 eV pump with 2.5~mJ/cm$^2$ fluence (D). (\textbf{C}, \textbf{E}) Simulated maps under the conditions of (B, D), respectively (see \cite{SM} for details).
    }
    \label{fig:diffmaps_exp_sim}
\end{figure*}

\noindent
Figure~\ref{fig:diffmaps_exp_sim} compares the variation in the loss function at the two different excitation energies (1.55~eV and 3.1~eV). As discussed above, the size of the photoexcited pockets changes as a function of photon energy: the 3.1~eV pump excitation induces larger e-h pockets in momentum space and, thus, requires phonons with smaller momentum $q$ for intervalley scattering, leading to a more isotropic phonon distribution in momentum space.
The energy-momentum maps in panels Figure~\ref{fig:diffmaps_exp_sim}B-E represent experimental and theoretical pump-induced EELS changes as a function of momentum and energy transfer, showing regions of alternating sign that can be attributed to plasmon shits. To visualize this effect, we plot in Figure~\ref{fig:diffmaps_exp_sim}A a characteristic EELS spectrum for undeflected electrons (i.e., at the $\Gamma_{000}$ point) acquired from unperturbed graphite, where we observe two  peaks corresponding to the in-plane and bulk plasmons described above; we then assume an excited spectrum with both plasmons shifted to higher energy and compute the difference with respect to the unshifted spectrum. The energy shift results in positive (red) and negative (blue) differential signals forming a pattern analogous to those in panels B-E as a function of momentum.

\noindent
When using NIR pulses (1.55~eV, Figure~\ref{fig:diffmaps_exp_sim}B), we find that, close to the $\Gamma$ point, the intensity of the in-plane $\pi$ plasmon decreases and the bulk $\sigma$+$\pi$ plasmon moves toward higher energies. We observe the opposite trend at larger momentum transfer, with an increase of the $\pi$ plasmon and a more pronounced redshift of the bulk plasmon. 
The EELS response for 3.1~eV photoexcitation, illustrated in Figure~\ref{fig:diffmaps_exp_sim}D, instead shows a different behavior, namely: the $\pi$ plasmon shifts towards higher energies at small momenta ($q\le0.5\AA^{-1}$), while the bulk plasmon intensity of the $\sigma+\pi$ peak decreases, indicating a broadening without any clear energy shift. 

\noindent
To understand the microscopic origin of the EELS response in out-of-equilibrium graphite, we perform {\it ab initio} simulations for both equilibrium and photoexcitation conditions. We identify two contributions to the excited state: the effect of thermal expansion --as the plasmonic response largely depends on the interlayer spacing~\cite{Marinopoulos2004}-- and the variation in the population of strongly coupled optical phonons.  For the former, we use in-plane and out-of-plane thermal expansion coefficients taken from the literature~\cite{MM05,P93}, which directly modify the lattice constants depending on an effective temperature $T_{L}$, and this in turn results in a modified EELS probability $\mathrm{EELS}_{L}(T_{L})$. For optical-phonon coupling, we simulate the effect of a population of optical $E_{2g}$ and $A_1'$ phonons at the $\Gamma$ and the $K$ points, respectively; more precisely, we calculate EELS spectra for distorted unit cells corresponding to different atomic displacements along these modes; the EELS probabilities $\mathrm{EELS}_{A_1'}(T_{A_1'})$ and $\mathrm{EELS}_{E_{2g}}(T_{E_{2g}})$ are then computed as the average weighted by the probability densities associated with each specific value of the displacement (i.e., the amplitude distribution associated to ladder modes in a harmonic oscillator, weighted by the Bose-Einstein distributions at the phonon temperatures $T_{A_1'}$ and $T_{E_{2g}}$, respectively). The simulated EELS probability is finally obtained as $\mathrm{EELS}_{\mathrm{sim}}= w_{A_1'}\mathrm{EELS}_{A_1'}(T_{A_1'})+w_{E_{2g}}\mathrm{EELS}_{E_{2g}}(T_{E_{2g}})+w_{L}\mathrm{EELS}_{L}(T_{L})-\mathrm{EELS}_{eq}$, with fitted weights $w_{A_1'}=w_{E_{2g}}=0.25$ and $w_{L}=0.5$. We refer to~\cite{SM} for further details, along with values of the fitted temperatures $T_{A_1'}$, $T_{E_{2g}}$, and $T_{L}$ at each of the two pumping photon energies (see Table S2).

The results of our simulations are shown in Figure~\ref{fig:diffmaps_exp_sim}C, E, normalized as a function of momentum analogously to the experimental ones. For 3.1~eV pumping, we observe a large degree of similarity between the experimental and simulated maps. In particular, the experimental dynamics of both plasmons are well reproduced from small momentum up to $q\simeq$ 1~\AA$^{-1}$ by considering an out-of-equilibrium populations of the strongly coupled optical phonons. In addition, the experimental amplitude variation matches the simulations quantitatively. However, for NIR excitation, our simulations can only partially replicate the experimental observations. While the response at small momenta can be understood as arising from phonon contributions, the large variations of the plasmon at higher momentum transfer can only be reproduced by introducing an expansion of the interlayer spacing (the lattice term in the EELS simulations).

\section*{Discussion and outlook}

Collective modes in materials emerge from the coherent excitation of structural, electronic, or magnetic degrees of freedom. They span a wide range of energies extending  from $\mu$eV's in spin waves to meV's in phonons, or even to eV's in plasmons and excitons. In addition, these modes can couple together {\it via} specific interaction processes, giving rise to exotic many-body objects such as charged phonons~\cite{Kuzmenko2009} or phonon-assisted bimagnons~\cite{Perkins1993, Lorenzana1995}. The capability to investigate the microscopic details associated with the interaction among collective modes and with single-particle excitations in materials is key to understand the scattering mechanisms that govern the energy dissipation ensuing an out-of-equilibrium excitation, and more generally, the dynamics associated with energy pathways in excited materials. To this end, we have shown that one needs to combine spectroscopic and momentum-resolved information on the collective mode dynamics. 
Ultrafast momentum-resolved EELS in a UTEM is a particularly powerful approach, as it grants continuous access to several momentum channels at once, in contrast to x-ray scattering or Raman spectroscopy, which can only measure specific momentum transfers, while being able to monitor a broad spectral range encompassing several electronic excitations of interest. Moreover, its combination with high signal-to-noise ratio UED complements this rich information by directly tracking across momentum space the diffuse scattering dynamics of low energy collective modes, that are not observable by tr-q-EELS because of its limited energy resolution. 

\noindent
In graphite, the photoexcited e-h pockets define the resulting inter- and intra-valley scattering dynamics. Merging together the results from UED and tr-q-EELS, we observe that for different excitation energies the preferential scattering channel varies and that the out-of-equilibrium phonon population renormalizes collective oscillations of the charge distribution. Our theoretical model accounts for the interactions between plasmons and phonons. Namely, we can quantitatively understand the response after a visible laser pulse, but our simulations fall short on explaining the plasmon dynamics with NIR photoexcitation. 
Based on current theoretical knowledge we can only speculate on the causes of such an evident shift in plasmon energy that cannot be accounted for by phonon-mediated scattering. In our model, we did not consider the role of the high electronic temperature that can contribute to changes in the plasmon response \cite{Stange2015}.

\noindent
We demonstrate that our comprehensive approach enables the resolution of microscopic details of collective modes in condensed materials, revealing for the first time that the scattering mechanism of charge carriers can be controlled by tuning the pump energy, resulting in the steering of the collective plasmonic response as a function of time and momentum on picosecond timescales.
Leveraging the high spatial resolution of transmission electron microscopes, this method introduces a novel tool for directly investigating the properties of emerging quasi-particles within nanoscale 2D platforms. Simultaneously, this protocol can be applied to other materials such as cuprates to unfold the interplay between high- and low-energy collectives modes~\cite{Levallois2016, Mitrano2018, Barantani2022} across the BZ. More generally, revealing the momentum dependence of collective excitations in photo-induced exotic phases~\cite{Fausti2011, Li2019, Nova2018} can help to shed light on their origin and unveil microscopic interactions in out-of-equilibrium strongly correlated states of matter.


\clearpage
\end{document}